\documentclass[sigconf]{acmart}

\AtBeginDocument{%
  \providecommand\BibTeX{{%
    \normalfont B\kern-0.5em{\scshape i\kern-0.25em b}\kern-0.8em\TeX}}}

\acmConference[SIGIR '21]{SIGIR '21:ACM SIGIR Conference on Research and Development in Information Retrieval}{}{}
\acmBooktitle{SIGIR '21:ACM SIGIR Conference on Research and Development in Information Retrieval, Under Review}
\acmPrice{15.00}
\acmISBN{978-1-4503-XXXX-X/18/06}
\usepackage{ulem}

\usepackage{paralist}
\usepackage{xcolor, soul}
\usepackage{multirow}
\usepackage{ulem}
\usepackage{amsthm,amsmath,amsfonts,bm,xspace}
\usepackage{bbm}
\usepackage{color}

\def\eqref#1{(\ref{#1})}

\def\1{\bm{1}}

\def\rmG{{\mathbf{G}}}

\def\rmP{{\mathbf{P}}}
\def\rmQ{{\mathbf{Q}}}

\def\vone{{\bm{1}}}

\def\vc{{\bm{c}}}

\def\vr{{\bm{r}}}

\def\vu{{\bm{u}}}
\def\vv{{\bm{v}}}

\def\mK{{\bm{K}}}

\def\mM{{\bm{M}}}

\def\mT{{\bm{T}}}

\DeclareMathAlphabet{\mathsfit}{\encodingdefault}{\sfdefault}{m}{sl}
\SetMathAlphabet{\mathsfit}{bold}{\encodingdefault}{\sfdefault}{bx}{n}

\definecolor{OliveGreen}{rgb}{0,0.8,0.1}
\newcommand{\model}{\mbox{\textsc{Rudder}}\xspace}
\begin{document}

\title{\model: A Cross Lingual Video and Text Retrieval Dataset}

\author{Jayaprakash A}
\authornote{Both authors contributed equally to this research.}
\author{Abhishek}
\authornotemark[1]
\affiliation{%
  \institution{Indian Institute of Technology, Bombay}
  \country{India}
}
\author{Rishabh Dabral}
\affiliation{%
  \institution{Indian Institute of Technology, Bombay}
  \country{India}
}

\author{Ganesh Ramakrishnan}
\affiliation{%
  \institution{Indian Institute of Technology, Bombay}
\country{India}
  }
\author{Preethi Jyothi}
\affiliation{%
 \institution{Indian Institute of Technology, Bombay}
\country{India}
 }

\renewcommand{\shortauthors}{Jayaprakash and Abhishek, et al.}
\begin{abstract}
	Video retrieval using natural language queries requires learning semantically meaningful joint embeddings between the text and the audio-visual input. Often, such joint embeddings are learnt using pairwise (or triplet) contrastive loss objectives which cannot give enough attention to ‘difficult-to-retrieve’ samples during training. This problem is especially pronounced in data-scarce settings where the data is relatively small (10\% of the large scale MSR-VTT) to cover the rather complex audio-visual embedding space. In this context, we introduce \model\ - a multilingual video-text retrieval dataset that includes audio and textual captions in Marathi, Hindi, Tamil, Kannada, Malayalam and Telugu. Furthermore, we propose to compensate for data scarcity by using domain knowledge to augment supervision. To this end, in addition to the conventional three samples of a triplet (anchor, positive, and negative), we introduce a fourth term - a partial - to define a differential margin based partial-order loss. The partials are heuristically sampled such that they semantically lie in the overlap zone between the positives and the negatives, thereby resulting in broader embedding coverage. Our proposals consistently outperform the conventional max-margin and triplet losses and improve the state-of-the-art on MSR-VTT and DiDeMO datasets. We report benchmark results on \model\ while also observing significant gains using the proposed partial order loss, especially when the language specific retrieval models are jointly trained by availing the cross-lingual alignment across the language-specific datasets.
\end{abstract}

\maketitle

\keywords{Video-Text Retrieval \and Multi-modal Learning \and Multi-lingual dataset}

\section{Introduction} \label{sec:intro} 
Learning low-dimensional, semantically rich representations of audio, visual and textual data (associated with videos) is a fundamental problem in multimodal learning~\cite{Mithun2018JointEW,Liu2019a,im3}. Such multimodal representations play a critical role in facilitating a multitude of visual analytics tasks such as Video Retrieval~\cite{Antoine_icc19}, Visual Question Answering (VQA), video summarization \textit{etc}.

Learning such compact representations is already a challenging task owing to the high dimensionality and multi-modality of sensory data. This difficulty is further exacerbated when the available data is limited and biased. With these challenges in focus, this paper introduces a new cross-lingual video dataset, \model\footnote{\model\ stands for c{\bf\uline{R}}oss ling{\bf \uline{U}}al vi{\bf \uline{D}}eo  an{\bf \uline{D}} t{\bf \uline{E}}xt {\bf \uline{R}}etrieval.},  with audio and textual descriptions in multiple languages for information retrieval tasks.  We also propose a novel approach to learning joint-embeddings between videos (audio-visual) and text while specifically being cognizant of its effectiveness for data-scarce settings.

Existing multi-lingual video datasets contain only bilingual textual descriptions~\cite{How2, VATEX} and may contain audio in at most one language~\cite{How2}. We present a new resource, \model\footnote{\url{https://rudder-2021.github.io/}}, that is truly multi-lingual with both, captions and narrations, in Marathi, Tamil, Telugu, Kannada, Malayalam and Hindi, thus facilitating training of cross-modal (audio/text/video) and cross-lingual embeddings. To the best of our knowledge, we are the first to propose a cross-linguo-modal audio-visual and text dataset.
We are also the first to create such a resource catering to Indian languages, which are extremely under-represented when considering multimodal resources.

We demonstrate the high impact of \model by using it to learn cross-linguo-modal embeddings for video-to-text and text-to-video retrieval tasks, across several languages. Specifically, we show that \model could be used for cross-linguo-modal tasks in which the audio track and textual captions may come from different languages but they project to the same point in the embedding space. As part of a competitive approach advancing state-of-the-art, we also introduce a novel partial-order contrastive loss that leverages augmented supervision to learn dense embeddings even for data-scarce cases. A typical approach to learning joint video-text embeddings is to use triplet/contrastive losses which prove insufficient in providing dense \textit{coverage} of the embedding space, particularly in data-scarce settings. We propose an alternative solution in the form of a differential margin based \textit{partial order contrastive loss} that facilitates augmenting the supervision using domain knowledge. 
We argue that the triplet formulation, while crucial, is not sufficient. To further ease training, we propose to learn the embeddings using a quadruplet of \{anchor, positive, \textit{partial}, negative\} instances. The additional \textit{partial} samples can be also sampled using meaningful heuristics based on domain knowledge, such that the semantic interpretation of each \textit{partial} sample falls somewhere in between the positive and negative samples. We demonstrate that these heuristics can be almost as effective as manually identified \textit{partial} samples that we also release as part of the dataset (\model) associated with this paper.
\begin{figure}    
	\centering
	\includegraphics[width=0.47\textwidth]{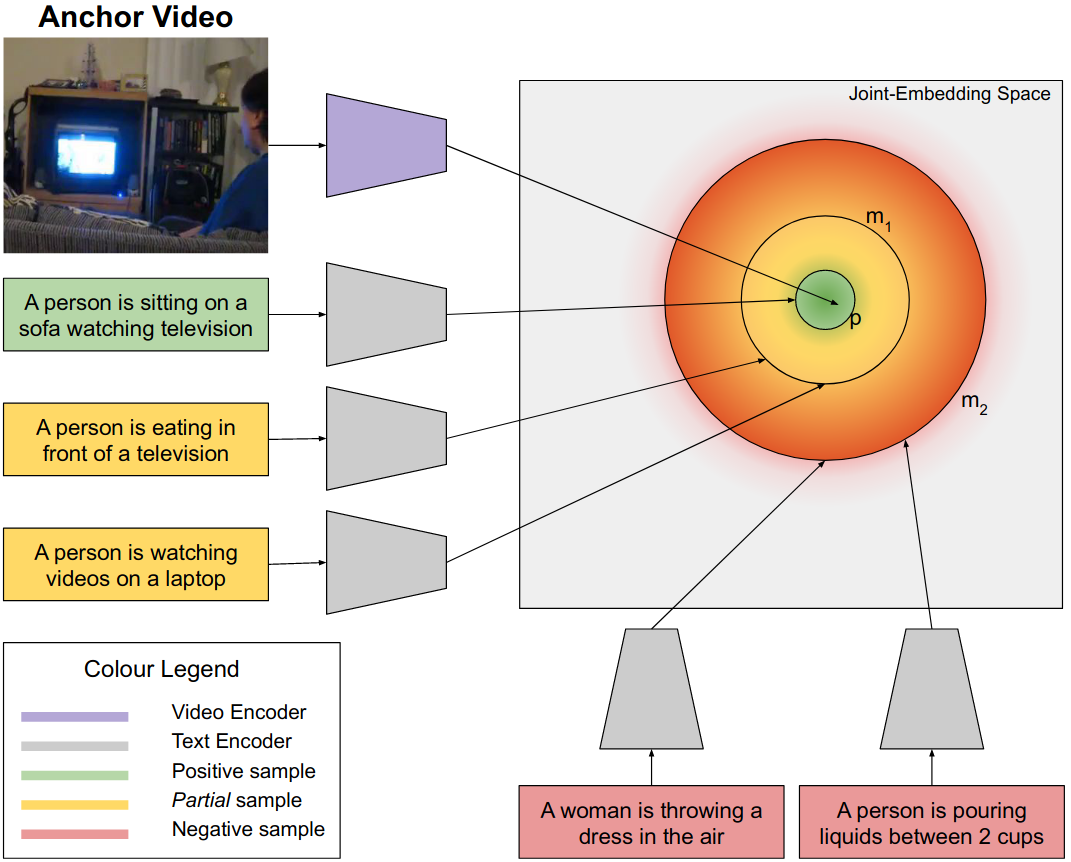}
	\caption{Figure illustrating the different roles played by \textit{positive}, \textit{partially} overlapping and \textit{negative} samples in our proposed simple differential margin based partial-order loss. Intuitively, we would like distances between the partially similar pairs to lie in the dark orange zone between the positive and the negative instances, \textit{i.e.},  beyond the $m_1$ margin but within $m_2$. Distances between unrelated pairs should be pushed to the grey region beyond $m_2$, whereas distances between related pairs should be in the green region within the  margin $p$.}
	\label{fig:Dataset}
\end{figure}
We illustrate the intuitiveness of the proposal with the help of an example in Figure~\ref{fig:Dataset}, showing a frame from a video of a man watching television. While the \textit{positive} sample,
\sethlcolor{green}
\hl{ `A person sitting on a sofa watching television'} and the \textit{negative} sample,  
\sethlcolor{pink}
\hl{ `A woman is throwing a dress in the air'} are self-explanatory, the \textit{partially} relevant sample, 
\sethlcolor{yellow}
\hl{ 'A person is watching videos on a laptop'} is neither a strictly positive nor a strictly negative instance. Though the objects are different, the sentences still capture the act of a person watching a (similar-looking) object. We postulate that placing such sentences at an appropriate distance from the anchor can crucially contribute towards coverage within the sparse latent space shared between the videos and the text. 
When training with partial instances, we differentiate them from the anchors using a partial-margin value that lies in between the margin values of positive and negative samples. 

We evaluate the proposed partial-order loss on the \model dataset and demonstrate its effectiveness with respect to strong baselines such as the Max-Margin contrastive loss, the Optimal Transport loss proposed in~\cite{xu2019learning}, \textit{etc.}
We also validate the similar effectiveness of the partial-order loss on existing benchmark datasets such as MSR-VTT~\cite{xu2016msr}, Charades~\cite{gao2017tall}, DiDeMo~\cite{hendricks-etal-2018-localizing}, \textit{etc.}

In summary, the contributions of this paper are as follows: 
\begin{itemize}
	\item 	We release a newly annotated cross-lingual video dataset, \model, to test our (and relevant) methods on data-scarce settings. The dataset (described in Section~\ref{sec:rudder}) includes audio-visual footages in Marathi, Hindi, English, Tamil, Kannada, Malayalam and Telugu. A significant percentage (75\%) of the videos are the same, and for these videos, the audio tracks are largely aligned
	      across the languages. Furthermore, in addition to providing multi-lingual audio for each video, we also provide textual captions in corresponding languages along with the associated start and end timings.
	\item We release a new challenge based on \model that we hope will spur more interest in designing new  cross-linguo-modal techniques.
	\item We propose a new quadruplet-based partial order loss (in Section~\ref{sec:poloss}) to improve the coverage of video-text retrieval models. The proposed intuitive loss is a competitive approach that advances state-of-the-art by taking into account augmented supervision along with the already available ground truth relevance. In this loss, we tune margins (such as $m1$ and $m2$ in Figure~\ref{fig:Dataset}) that help differentiate positives from the partial and the partial from the negatives. 
\end{itemize}
\section{Related Work}
\paragraph{MultiLingual Visual Datasets:}
While vision and language have been used in conjunction in several datasets~\cite{xu2016msr, hendricks-etal-2018-localizing, gao2017tall}, multi-lingual visual datasets have continued to be under-explored. \cite{Gao2015AreYT} proposed a multi-lingual image dataset, FM-IQA, for image question answering in  Chinese and English. \cite{How2} propose a Portuguese/English dataset with audio, video and textual captions for multi-modal learning.  \cite{VATEX} introduced a bi-lingual video-text dataset, VaTex for Chinese and English. However, none of the above mentioned datasets goes beyond two languages to be truly multi-lingual. In comparison, our proposed \model dataset is a multi-lingual audio-visual and textual dataset with audio narrations and textual captions in six languages.

\paragraph{Learning Representations for Video-Text Retrieval:}
Several of the existing approaches for video-text retrieval~\cite{meera2019action,Michael2019fine,Mithun2018JointEW} are based on image-text embedding methods by design, and typically focus on single visual frames.

\cite{Michael2019fine} focus on learning separate embedding spaces for each POS tag, using triplet losses wherein the positive examples were constructed by finding relevant entities that shared common nouns and verbs. We borrow inspiration from this work while designing our heuristics.
Collaborative Experts~\cite{Liu2019a} is proposed as a network to effectively fuse the action, object, place, text and audio features extracted from state-of-the-art models (experts) with the reasoning that extracting supervision from multiple experts could help compensate for the lack of sufficient data. Our proposed loss function is used within the collaborative experts framework to further help improve coverage, specifically in low-resource settings.

\paragraph{Loss Functions for Cross-Modal Learning:} Several max-margin based ranking loss functions have been proposed for learning joint embeddings. The triplet loss~\cite{chechik2010large,facenet} and bidirectional max-margin ranking loss~\cite{socher2014grounded} have been popularly used to learn cross-modal embeddings. \cite{xu2019learning} propose a new approach to learn distance metrics via optimal transport programming for batches of samples and report faster convergence rates without adversely affecting performance. 

Significant effort has also gone into better sampling methods for training ranking losses. FaceNet~\cite{facenet} uses semi-hard negatives within the triplet-loss framework. Notably, \cite{wu2017iccv} propose a distance-based sampling method to choose stable and informative samples that can be trained with a simple max-margin loss. Likewise, \cite{wang_neighborhood} propose to consider samples in the neighborhood of positives. Our partial ordering based loss formulation subsumes~\cite{wang_neighborhood} in the sense that we also consider the extended neighborhood between the positives and negatives. The approach closest to our partial-order based loss, however, is the work of~\cite{Karaman_2019_IEEE}, who use the knowledge of category hierarchy to label samples as positives or semi-positives. In addition to differing from them in our loss formulation, it is worth noting that unlike categorical data, establishing hierarchies in textual queries is not straightforward.
\section{Background}
Embedding learning aims to learn semantically-aware and  compact fixed length representation vectors from input samples. A kernel function $f(\cdot; \theta): X \rightarrow R^d$ takes input $x \in X$ and generates a feature representation or an embedding.
$f(\cdot; \theta)$ is usually defined by a deep neural network, parameterized by $\theta$. Embedding learning optimizes a discriminative loss objective to minimize intra-class distances while maximizing inter-class distances. For
example, the contrastive loss in the seminal work introducing Siamese networks \cite{hadsell2006dimensionality} takes pairs of samples as input and trains two identical networks to learn a deep metric $M$ by minimising the following loss functional,
\begin{equation}
	\mathcal{L}(x_i, x_j ; f) = y_{ij}M_{ij} +(1 - y_{ij}) \max\{0, \epsilon - M_{ij}\}
\end{equation}
where the label $y_{ij} \in \{0, 1\}$ indicates whether a pair of $(x_i, x_j )$ is from the same class or not and $M_{ij}$ denotes the distance between $x_i$ and $x_j$. The margin parameter $\epsilon$ imposes a threshold of the distances among dissimilar samples. Triplet loss \cite{chechik2010large} is conceptually similar to the contrastive loss, and extends pairs of samples to triplets. Given a query $x_i$ with a sample $x_j$ similar to $x_i$ and a dissimilar sample $x_k$,
the triplet loss function is formulated as,
\begin{equation}
	\mathcal{L}(x_i, x_j , x_k; f) = \max\{0,M_{ij} - M_{ik} + \epsilon\}.
\end{equation}
Intuitively, it encourages the distance between the dissimilar pair $M_{ik}$ to be larger than the distance between the similar pair $M_{ij}$ by at least a margin $\epsilon$.

Optimal transport distances, also known as Wasserstein distances~\cite{vallender1974calculation} or Earth Mover’s distances~\cite{rubner2000earth}, define a distance between two probability distributions according to principles from optimal transport theory~\cite{villani2008optimal}. 
Formally, let $\vr$ and $\vc$ be $n$-dimensional probability measures. The set of transportation plans between probability distributions $\vr$ and $\vc$ is defined as $U(\vr, \vc) := \{\mT \in R^{n\times n}_+ |\mT \cdot \vone = \vr, \mT^T \cdot \vone = \vc\}$, where \textbf{1} is an all ones vector. The set of transportation plans $U(\vr, \vc)$ contains all non-negative $n\times n$ elements with rows and columns summing to $\vr$ and $\vc$, respectively. Give an $n \times n$ ground distance matrix $\mM$, the cost of mapping $\vr$ to $\vc$ using a transport matrix $\mT$ is quantified as $\langle \mT ,\mM\rangle$, where $\langle., .\rangle$ denotes the Frobenius norm (dot product). The resulting optimisation problem is then,
\begin{equation}
	D_{\mM}(\vr, \vc) := \min_{\mT \in U(\vr,\vc)} 
	\langle \mT ,\mM\rangle 
	\label{eqn:emd}
\end{equation}
which is called an optimal transport problem between $\vr$ and $\vc$ given ground cost $\mM$. The optimal transport distance $D_{\mM}(\vr, \vc)$ measures the cheapest way to transport the mass in probability measure $r$ to match that in $c$. Optimal transport distances define a more powerful
cross-bin metric to measure probabilities compared with
some commonly used bin-to-bin metrics, e.g., Euclidean,
Hellinger, and Kullback-Leibler divergences. However, the cost of computing $D_M$ is at least $O(n^3log(n))$ when comparing two $n$-dimensional probability distributions in a general metric space. To alleviate it, \cite{cuturi2013sinkhorn} formulated a regularized transport problem by adding an entropy regularizer to Eqn.~\eqref{eqn:emd}. This makes the objective function strictly convex and allows it to be solved efficiently. Particularly, given a transport matrix $\mT$, the entropy of $\mT$ can be computed as $h(\mT) = \displaystyle -\sum_{ij} \mT_{ij} \log \mT_{ij}$. For any $\lambda > 0$, the regularized transport problem can be defined as
\begin{equation}
	D^{\lambda}_{\mM}(\vr, \vc) := \min_{\mT \in U(\vr,\vc)} \langle \mT, \mM\rangle - \frac{1}{\lambda} h(T)
\end{equation}

\noindent where the larger $\lambda$ is, the closer this relaxation $D^{\lambda}_{\mM}(\vr, \vc)$ is to the original $D_{\mM}(\vr, \vc)$. \cite{cuturi2013sinkhorn} also proposed the Sinkhorn’s algorithm to solve Eqn. (4) for the optimal transport $\mT^\ast$. Specifically, let the matrix $\mK = \exp(-\lambda \mM)$ and solve it for the scaling vectors $\vu$ and $\vv$ to a fixed-point by computing $\vu = \vr./\mK \vv$, $\vv = \vc./\mK^T u$ in an alternating way. This yields the optimal transportation plan $T^{\ast} = \text{diag}(\vu)\mK \text{diag}(\vv)$, which can be solved significantly faster ($O(n^2)$) than the original transport problem.
\section{The \model\ Dataset}
\label{sec:rudder}

We propose a cross-lingual video and text dataset for retrieval tasks on a set of languages that include Marathi, Hindi, English, Telugu and Tamil. 
The dataset consists of video tutorials for making scientific toys from waste material with audio narrations in multiple languages.\footnote{We downloaded these videos from \url{http://www.arvindguptatoys.com/toys-from-trash.php} and obtained consent from the content creator to use the videos for research.} 
Each video is associated with an audio track that consists of a descriptive narration of the process of making those toys.%
\footnote{Please refer to \url{https://rudder-2021.github.io/tutorial.html} for additional details about the dataset.}
We also provide transcriptions of the audio tracks, so that they serve as textual captions/descriptions for the videos. Since the videos are instructional in nature, they support a meaningful association of objects with the underlying toy-making process. To the best of our knowledge, we are the first to release a cross-lingual video dataset with multi-lingual audio and textual tracks. 
\paragraph{Annotation Process}
Each video in the dataset originally came with audio narrations in Marathi, Hindi,  English, Tamil, Telugu, Kannada and Malayalam. Of these, the audio tracks for Marathi and Hindi were manually transcribed by native speakers to help construct the text captions. For the remaining languages, we construct the captions by translating the Marathi captions into those respective languages using the Google Translate API. While we could have used an ASR system to transcribe the audio of the remaining languages, we found the ASR based transcriptions to be severely lacking in quality due to poor support for these specific languages. \looseness-1
\begin{table}[!h]
\centering
\resizebox{\linewidth}{!}{
		\begin{tabular}{|c|c|c|c|}
			\hline
			\multirow{2}{*}{\textbf{Language}} & \textbf{Mean \& Std dev}      & \multirow{2}{*}{\textbf{Language}} & \textbf{Mean \& Std dev}      \\
			                                   & \textbf{Speech Length (secs)} &                                    & \textbf{Speech Length (secs)} \\ \hline
			Hindi                              & 6.63 $\pm$ 3.91 (92.3\%)      & Kannada                            & 6.05 $\pm$  3.70 (77.2\%)     \\ \hline
			Malayalam                          & 5.81 $\pm$ 3.70 (77.1\%)      & Marathi                            & 6.40 $\pm$  3.79 (87.5\%)     \\ \hline
			Tamil                              & 5.95 $\pm$ 3.82 (83.4\%)      & Telugu                             & 6.03 $\pm$ 3.80 (80.5\%)      \\ \hline
		\end{tabular}
	}
	\caption{Mean and standard deviation of the length of speech (in seconds) across languages in \model. The numbers in parenthesis show the percentage of human speech in the entire audio clip}
	\vspace{-10mm}
\end{table}
\paragraph{Dataset Statistics: } The dataset consists of 3272 videos overall.  We follow a 70:5:25 percentage split for the train, validation and test sets respectively. Specifically, the training set consists of 2287 videos, 166 videos are identified for validation and 819 videos form part of the test data. 
Table 1 lists the amount of human speech available in our dataset to quantify the richness of our audio files. The amount of human speech was calculated using an off-the-shelf voice activity detection model as in~\cite{VAD}. 
The average length of a video is 8 secs and each video is associated with a sentence describing it. Of the 3272 videos, 675 videos (roughly 21\%) include audio narrations in all languages. 54\% videos come with Hindi and Marathi narrations, 27\% with Tamil and Marathi, 53\% with Kanadda and Marathi, 33\% with Malayalam and Marathi, and 23\% with Telugu and Marathi.  %
\subsection{Annotations for Augmented Supervision \label{sec:augsuper}}
As motivated in Section~\ref{sec:intro} and illustrated in Figure~\ref{fig:Dataset}, we propose augmented supervision for guiding a loss (such as the partial ordering based loss proposed in Section \ref{sec:poloss}).  
To enable a sanity check on the process of heuristically generating augmented supervision, we also provide manually generated augmented supervision. 
For each query caption in the dataset, these manual annotations consist of partial/positive/negative labels assigned to video segments corresponding to that caption that are obtained using the following process.
For each query/anchor caption in the dataset, we first find the top $K=10$ `similar' captions based on sentence similarity scores while ensuring that these `similar' captions do not belong to the same video as that of the query caption. The annotators are then asked to assign one out of the following three relevance judgements to each pair of similar captions/video segments. We will now illustrate the relevance judgements in Figure~\ref{fig:Dataset} with respect to the example query caption `\textit{A person is sitting on a sofa watching television}':  

(i) \textbf{Positive}- the relevance judgement when both the captions/video segments in the pair share an almost similar amount of information. In Figure~\ref{fig:Dataset}, the following caption (and its corresponding video segment) is labeled \textbf{positive} with respect to the query caption:
\sethlcolor{green}
\hl{`A person is sitting on a sofa watching television'}  
(ii) \textbf{Partial}- the relevance judgement when both the captions/video segments in the pair share somewhat similar information. In Figure~\ref{fig:Dataset}, the following caption (and its corresponding video segment) is one of the two labeled \textbf{partial} with respect to the query caption:
\sethlcolor{yellow}

\hl{`A person is eating in front of a television'}.
(iii) \textbf{Negative}- the relevance judgement when both the captions/video segments in the pair are irrelevant to each other. In Figure~\ref{fig:Dataset}, the following caption (and its corresponding video segment) is one of the two labeled \textbf{negative} with respect to the query caption: 
\sethlcolor{pink}
\hl{`A woman is throwing a dress in the air'}.

We provide two sets of relevance judgements with the dataset, one that is manually determined (using guidelines described in detail on our website) and the other based on an automated, domain-knowledge driven \textit{Noun-Verb heuristic}.  The annotations were performed on approximately 32700 sentence pairs. While the manual annotation task was divided across three annotators, a separate annotator verified all the annotations. All this amounted to around 200 hours of annotation work cumulatively.
\paragraph{Noun-Verb Heuristic:} For a given query caption, this heuristic labels another caption as \textit{partially} relevant if either the two captions have exactly the same set of nouns but differ only in their verbs (\textit{e.g.}, `person eating cake; person sitting next to a cake) or they differ in their nouns but share the same verb(s) (\textit{e.g.}, `person eating cake'; `person eating cereals'). A caption is labelled  \textit{positive} with respect to the query if both share the same nouns and verbs, whereas they are labeled \textit{negative} if they share neither the nouns or verbs. 

\subsection{The \model\ Challenge}
We release the \model\ dataset along with an associated cross-lingual cross-modal retrieval challenge\footnote{\url{https://rudder-2021.github.io/challenge.html}}. The challenge requires performing video-to-text and text-to-video retrieval based on queries in multiple languages. The submissions are evaluated on 3 metrics, {viz.}, Recall@K (with K = 1, 5 , 10, and 50) Median Rank, and Mean Rank. The challenge is organized under various tracks, with each track consisting of text from either Hindi or Marathi and audio from either of the four audio languages (Hindi, Marathi, Tamil, Telugu). Currently, we have eight tasks of the form lang1-text+lang2-audio, each of which denotes a video-text retrieval task using lang1 for caption and lang2 for audio. For the submission format, please refer to the challenge website.

We envisage several other potential uses/impacts of the \model\ challenge. These include training/evaluation for 
\begin{enumerate}
	\item Cross-lingual video-caption retrieval. 
	\item Multi lingual video recommendation: The manually annotated \textit{positive} and \textit{partial} tags could serve as video recommendations, in a setting in which the user has only the videos and no textual meta-data.  
	\item Multilingual and Cross lingual video summarisation, aimed at describing a video in various languages, via captions. 
	\item Video-guided machine translation, involving translation of a source language description into the target language using the video information as additional spatio-temporal context.
\end{enumerate} 
\section{A Competitive Approach Advancing State-of-the-Art}\label{sec:poloss} %
A typical approach to learning joint video-text embeddings involves projecting the video and text inputs onto a common latent space such that semantically similar video-text pairs are in the same neighbourhood whereas dissimilar video-text pairs are pushed far apart. This projection is achieved by training video and text encoders using triplet ranking losses which operate on triplets of $<$anchor, positive, negative$>$ samples~\cite{chechik2010large}. Here, the anchor can be a video embedding and the positive and negative parts can correspond to instances with textual embeddings or vice-versa.

While this training regime is suitable for resource-rich languages with availability of large audio-visual and textual datasets, the performance quickly deteriorates when applied to data-scarce settings due to two main reasons. Firstly, the scarcity of data weakens the video (and textual) encoders which do not \textit{cover} the shared embedding space densely enough, thereby restricting their ability to meaningfully project to the shared space. Secondly, data scarcity inevitably prevents the encoders from generalizing to out-of-distribution text or video queries. A natural solution to the problem would be to explicitly target the hard positive and hard negative instances in the data. This approach, however, has limited returns as it leads to noisy gradients, thereby leading to a collapsed model~\cite{wu2017iccv}. We propose a quadruplet-based partial-order augmented contrastive loss to learn joint video-text embeddings, which we described next.

Let $\mathcal{V} = \{v_1, v_2, \ldots, v_{\vert \mathcal{V} \vert}\}$ and $\mathcal{T} = \{t_1, t_2, \dots, t_{\vert \mathcal{T} \vert}\}$ denote a set of videos and captions. We define $f(\cdot)$ and $g(\cdot)$ as video and caption encoders, respectively, both of which embed an $i^{th}$ video, $v_i$, and a $j^{th}$ caption, $t_j$, into an $N$-dimensional space. Let ${d}_{i,j} = \mathrm{dist}(f(v_i),g(t_j))$ be the distance between the embeddings corresponding to video $v_i$ and caption $t_j$.

The standard \textit{bidirectional max-margin ranking loss}~\cite{socher2014grounded} can be written as:
\begin{equation}
	\begin{aligned}
		\mathcal{L} =  \sum_{i,j \ne i }[m + d_{i,i} - d_{i,j}]_+ +  [m + d_{i,i}- d_{j,i}]_+. 
		\label{eq:maxMargin}                                                                   
	\end{aligned}
\end{equation}
Here, $m$ is a tunable margin hyperparameter that 
lower-bounds the distance between negative pairs and $[.]_+$ represents the $\max(., 0)$ function. 


\subsection{Optimal Transport Loss}
In addition to subjecting the pairs of positive, negatives and partials to margin based ranking losses, we also test the efficacy of heuristic-based augmented supervision on Optimal Transport based loss function as proposed by~\cite{xu2019learning}. 
As an alternative to the differential margins, one could  differentialy weigh the difficult pairs and contrast them against  the easier ones. Toward this, we consider a batch-wise Optimal Transport (OT) based loss, building upon the work of~\cite{xu2019learning}. 
The originally proposed loss defines ground distances, $G^+$ and $G^-$ for positive and negative pairs and uses them to estimate an optimal transport plan, $T^*$, that assigns higher importance to difficult pairs and vice-versa.  

\begin{align}
    \begin{split}
	G_{i,j}^+ ={}& e^{-\gamma \left([d_{i,j} - d_{i,i} - p]_+ + 	[d_{j,i} - d_{i,i}- p]_+\right) }, 
    \end{split}\\
    \begin{split}
	G_{i,j}^- ={} &e^{-\gamma([n -d_{i,j} + d_{i,i}]_+ + 	[n -d_{j,i}+ d_{i,i}]_+) } 
    \end{split} 
\end{align}
\begin{equation}
\begin{aligned}
T^* &=& \mathrm{arg\,min}_T\sum_{i,j} \bigg( \rmP_{i,j}(1-\rmQ_{i,j})T_{i,j}\rmG_{i,j}^+ + (1-\rmP_{i,j})\rmQ_{i,j}T_{i,j}\rmG_{i,j}^- \bigg) \nonumber
\end{aligned}
\end{equation}
where, 
$\rmP_{i,j}=1$ if $(i,j) \in S^+$, else 0. and
$\rmQ_{i,j}=1$ if $(i,j) \in S^-$, else 0.
\begin{equation}
	\mathcal{L}^{OT} = \sum_{i,j}T^*_{i,j}\mathcal{L}_{i,j}^{MM}
\end{equation} 

\subsection{Partial Order Contrastive Loss}
The bidirectional max-margin loss defined above separates the negative instances from positive instances by a fixed margin, $m$. Although this is regarded as one of the standard ways of learning cross-modal embeddings, we argue that in the absence of a large dataset, the loss results in a sparsely \textit{covered} embedding space, thereby hurting the generalizability of the embeddings. Furthermore, the pretrained embeddings (GLoVE, GPT, etc) for low-resource/underrepresented languages are often either weak or simply may not exist. 

In order to circumvent the challenges associated with a low-resource language, we propose the following quadruplet based mining setup. Given a batch of videos and their corresponding captions, for every anchor sample, we construct three sets of video-caption pairs, \textit{viz.}, (i) the positive ($S^+$), (ii) the negative ($S^-$) and (iii) the \textit{partial}-overlap ($S^{\sim}$). While the positive and negative pairs are chosen as in the bidirectional max-margin loss, we show we can make effective use of dataset-dependent heuristics to sample the \textit{partial} samples. Intuitively, the \textit{partial} samples are chosen such that they represent a degree of semantic overlap with the anchor (as illustrated in Figure~\ref{fig:Dataset}). We formally define the proposed novel Partial Order (PO) Contrastive Loss as:
\setlength{\belowdisplayskip}{0pt} \setlength{\belowdisplayshortskip}{0pt}
\setlength{\abovedisplayskip}{0pt} \setlength{\abovedisplayshortskip}{0pt}
\begin{eqnarray}
	\mathcal{L}^{PO} &=& \mathcal{L}^+ + \mathcal{L}^- + \mathcal{L}^{\sim} \nonumber \\
	\mathcal{L}^+ &=& \sum_{(i,j) \in S^+} [d_{i,j} - d_{i,i} - p]_+ + 	
	[d_{j,i} - d_{i,i} - p]_+ \nonumber \\
	\mathcal{L}^- &=& \sum_{(i,j) \in S^-}
	[n + d_{i,i} - d_{i,j}]_+ + 	
	[n + d_{i,i} - d_{j,i}]_+ \nonumber \\
	\mathcal{L}^{\sim} &=& \sum_{(i,j) \in S^{\sim}}\Big(
	[m_{1} + d_{i,i} - d_{i,j}]_+ + 
	[m_{1} + d_{i,i} - d_{j,i}]_+ \nonumber  \\ &+&
	[d_{i,j} - d_{i,i}  - m_{2}]_+ + 
	[d_{j,i} - d_{i,i}  - m_{2}]_+\Big)  
	\label{eq:partial:related}
	\label{eq:partilOrderLoss}
\end{eqnarray}

\noindent Here, $p$, $m_1, m_2$ and $n$ are tunable margin hyperparameters such that $p < m_1 < m_2 < n$. 
{As depicted in Figure~\ref{fig:Dataset}, we would like distances between the partially similar pairs to lie in the dark orange zone between the positives and the negatives, \textit{i.e.}, beyond the $m_1$ margin but within $m_2$. Distances between unrelated pairs should be pushed to the blue region beyond $m_2$, whereas distances between positive pairs should be in the green region within the margin $p$.}  
Note that the proposed setup is different from the similar-sounding paradigm of semi-hard negative mining~\cite{facenet}. While semi-hard negative mining is a useful tool to ease the training process, it does not deal with the issue of coverage over the embedding space.
\section{Experimental Setup}
While our primary experiments are conducted on the \model dataset, we also evaluate our proposed partial-order loss on three existing benchmark datasets:

\begin{enumerate}
	\item \textbf{Charades-STA}~\cite{gao2017tall} is a human-centric activity dataset containing one English caption per video.
	      We use a train/validation/test split of 3124/121/1015 clips respectively, while ensuring that each caption is uniquely mapped to one video.
	\item \textbf{MSR-VTT}~\cite{xu2016msr} is a relatively large video dataset consisting of 10K videos with multiple English captions per video. We follow a train-val-test split of 6512, 496, and 2990, respectively. 
	      To be consistent with other benchmarks and \model, and as in~\cite{Liu2019a}, we work with 1 caption per video.
	\item \textbf{DiDeMo:} Distinct Describable Moments (DiDeMo) ~\cite{hendricks-etal-2018-localizing} dataset consists of over 10K videos in diverse visual settings with pairs of localized video segments and referring expressions. Each video has around 3-5 descriptions aligned with time-stamps, which we concatenate into a single caption per video. We follow a train-val-test split of 8392, 1065 and 1004 videos respectively.
\end{enumerate}

\noindent \textbf{Evaluation Metrics:}
We use standard retrieval measures adopted from prior work~\cite{im1,Antoine_icc19,im3} and show results on both text-to-video (T2V) and video-to-text (V2T) retrieval. R@K (recall at rank K) for K=1,5,10,50 measures the percentage of the top-K retrieved text/video results, for a given video/text query, that match the ground-truth. Median Rank (MdR, lower is better), and Mean
Rank (MnR, lower is better)  compute the median and mean of the ground truth appearing in the ranking of the predictions, respectively. When computing video to-sentence measures for datasets containing multiple independent sentences per video, such as the \model\   dataset,
we follow the evaluation protocol used in prior work~\cite{8353472,2018arXiv180906181D} which corresponds to reporting the minimum rank among all valid text descriptions for a given video query. 

\noindent \textbf{Implementation Details:}
We adopt the state-of-the art collaborative experts (CE) framework  from~\cite{Liu2019a} as our base model.  
We choose four pretrained models to serve as experts and derived features from the scene ~\cite{scene,i3d}, audio (VGG ~\cite{vggish}), objects ~\cite{object_resnext,object_senet} and action ~\cite{action_r2p1d}, that are subsequently fed into the video encoder as inputs. The embeddings from these features are further aggregated within the video encoder using a collaborative gating mechanism, resulting in a fixed-width video embedding (denoted as $f(v)$). 

In order to construct textual embeddings (denoted by $g(t)$), the query sentences are converted into sequences of feature vectors using pretrained word-level embeddings like GloVE and FastText. These word-level embeddings are aggregated using NetVLAD~\cite{arandjelovic2016netvlad}. We subject the output of the aggregation step to the text encoding architecture proposed in~\cite{Antoine_icc19}, which projects text features to different subspaces, one for each expert. Each text feature is further refined by computing attention over aggregated text features and thereafter passed through a feedforward and a softmax layer to produce a scalar for each expert projection. Finally, each expert is scaled and concatenated to form a fixed-width text embedding.
 
We use PyTorch for our implementations. The margin values $p, m_1, m_2$ and $n$ are tuned on the validation sets. More details are provided in the extended version linked on our site.

\noindent \textbf{Baselines:} We compare the use of the partial order loss against the following baselines (i) CE trained with the standard max-margin loss (MM) (ii) an importance-driven distance metric via batch-wise Optimal Transport (OT) from Xu et al.~\cite{xu2019learning}. (iii) a Triplet loss (Triplet) baseline, (iv) HN baseline which refers to the hard negative mining strategy, where we pick only the hardest negative to train on and, finally, (v) a distance-weighted sampling based triplet loss (DW) from~\cite{wu2017iccv}, (vi) S2VT (Sequence to Sequence Video Text retrieval), which is a reimplementation of~\cite{venugopalan-etal-2015-translating},  (vii) FSE~\cite{Zhang_2018_ECCV}, which performs  Cross-Modal and Hierarchical Modeling of Video and Text and  
(viii) specifically on the MSRVTT and DiDeMo datasets, we also present the numbers as reported by \cite{Liu2019a} and refer to those as (MM~\cite{Liu2019a}). To the best of our knowledge, we are the first to evaluate the effectiveness of OT-based loss for retrieval tasks. 
\section{Results and Analysis}
We next describe experimental results that  validate the effect of using the proposed partial ordering based loss and compare them against the max-margin baseline as well as against the optimal transport based loss.   
\begin{table}[t!]
	\resizebox{\linewidth}{!}{
		\begin{tabular}{|c|c|c|c|c|c|c| }	
			\multicolumn{7}{c}{Text to Video Retrieval}\\
			\hline
			Loss    & R@1           & R@5            & R@10           & R@50           & MdR            & MnR             \\
			\hline\hline
			Triplet & 1.91          & 7.45           & 12.13          & 32.80          & 108.33         & 178.33          \\
			DW      & 2.16          & 7.45           & 12.13          & 33.41          & 110.33         & 184.11          \\
					
			MM      & 2.16          & 8.63           & 14.29          & 36.39          & 99.67          & 180.15          \\
			
			OT      & 2.85          & 9.40           & 14.86          & \textbf{37.53} & 99.33          & \textbf{175.23} \\
			
			\hline\hline
			
			PO      & \textbf{3.01} & \textbf{10.70} & \textbf{15.91} & 36.39          & \textbf{97.67} & 181.71          \\
			PO(M)   & \textbf{4.48} & \textbf{13.47} & \textbf{20.02} & \textbf{44.28} & \textbf{66.00} & \textbf{153.14} \\
			\hline
					
			\multicolumn{7}{c}{Video to Text Retrieval}\\ \hline
			Loss    & R@1           & R@5            & R@10           & R@50           & MdR            & MnR             \\
			\hline\hline
			Triplet & 1.99          & 7.33           & 12.09          & 33.66          & 107.00         & 173.32          \\
			DW      & 1.99          & 7.20           & 11.40          & 31.83          & 113.33         & 184.22          \\
					
			MM      & 2.12          & 9.08           & 13.84          & 36.18          & 96.00          & 174.87          \\
				
			OT      & 2.85          & 9.48           & 14.94          & 37.61          & 94.00          & \textbf{172.77} \\
			
			\hline	\hline
			
			PO      & \textbf{3.13} & \textbf{10.50} & \textbf{16.32} & \textbf{38.34} & \textbf{93.33} & 180.06          \\
			PO(M)   & \textbf{3.87} & \textbf{12.13} & \textbf{19.09} & \textbf{42.49} & \textbf{73.00} & \textbf{151.63} \\
			\hline
		\end{tabular}
	}
	\caption{\label{tab:malta_results}Results on human annotated \model\  dataset. PO and PO(M) denote the use of heuristically annotated and manually annotated partial samples for the partial order loss. DW refers to our implementation of Distance-Weighted sampling as in  \cite{wu2017iccv}.}
	
\end{table}
\subsection{Results on Synthetic Data} 
\begin{figure}[h]
	\centering
	\includegraphics[width=0.2\textwidth]{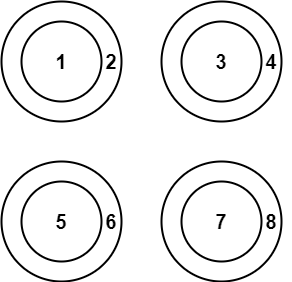}
	\caption{Synthetic dataset to illustrate the effectiveness of the partial order loss.} 
	\label{fig:synthetic_data}
\end{figure}

\paragraph{Synthetic Experiment.} Before presenting our results from experiments on existing datasets, we illustrate the importance of the proposed partial ordering based loss with the help of a synthetic task. This experiment also  highlights the relevance of the partial ordering based loss when training data is limited. Consider a simple dataset consisting of 8 classes as shown in Fig.~\ref{fig:synthetic_data}. Classes with odd-numbered labels correspond to the inner circles and classes with even-numbered labels correspond to the annular rings ({\em i.e.}, the outer circle minus the respective inner circle). The radii are chosen in such a way that all the classes have the same area.
While the model that gets trained by minimizing the max-margin based loss in Eqn~(\ref{eq:maxMargin}) uses only the ground truth for training, the partial ordering based loss based on Eqn~(\ref{eq:partilOrderLoss}) makes use of the following augmented supervision. For each query point that belongs to class $i \in \{1,3,5,7\}$, the samples coming from class $i+1$ should be within the margins $m_1$ and $m_2$ and the samples coming from any other classes other than $i$ and $i+1$ should be more than $m_2$ as described in Eqn ~(\ref{eq:partilOrderLoss})
Training points are sampled uniformly from all the 8 classes and the test data is a randomly sampled set of 20 points.Table~\ref{table:synthetic_results} shows  The numbers are obtained over an average of 5 retrieval  random train-test samples. Empirical results on the task of retrieval when performed using a model containing a single linear layer without any activation can be observed in . Note that  We observe that partial ordering  loss based formulation yields larger gains when the training datasets are smaller. The general observation is that the gains using partial order loss based model over the max-margin loss based model increase as the training data set size is decreased (until an unreasonably small training set size is reached). The explanation for the gain could be explained as follows: Consider a point from class 1. Then, according to partial order loss we can adjust points coming from class 2 to be within $m_1$ and $m_2$ and points coming from other classes to be at least $m_2$ apart. Whereas for the model using max margin loss it could be challenging for it to make points in class 2 to be at least as far away as any other points from other classes due to small amounts of training data.

Training points are sampled uniformly from all the 8 classes and the test data is a randomly sampled set of 20 points. Table~\ref{table:synthetic_results} shows the retrieval results when using a model with a single linear layer without any activation function for two different training set sizes (100 and 1000). All the numbers are averaged over five different draws of the train/test samples. We observe larger gains in performance with using our partial order (PO) loss compared to a max-margin (MM) loss, when the train set is smaller; these improvements are diminished with the larger training set. The gains from PO could be intuitively explained using the following example. Consider an anchor point from class 1. With PO, points from class 2 have to be within $m_1$ and $m_2$ away from the anchor point, and points from other classes should be at least $m_2$ apart. With MM, it could be challenging to push points in class 2 to be as far away as points from other classes due to small amounts of training data.

\begin{table}[t!]
	\resizebox{\linewidth}{!}{
		\begin{tabular}{|c|c|c|c|c|c|c| }	
			\multicolumn{7}{c}{Text to Video Retrieval}\\
			\hline
			Loss & R@1           & R@5            & R@10           & R@50           & MdR            & MnR             \\
			\hline\hline
			MM   & 2.16          & 8.63           & 14.29          & 36.39          & 99.67          & 180.15          \\
			
			OT   & 2.85          & 9.40           & 14.86          & \textbf{37.53} & 99.33          & \textbf{175.23} \\
			
			PO   & \textbf{3.01} & \textbf{10.70} & \textbf{15.91} & 36.39          & \textbf{97.67} & 181.71          \\
			\hline
			\hline
					
			\multicolumn{7}{c}{Video to Text Retrieval}\\ \hline
			Loss & R@1           & R@5            & R@10           & R@50           & MdR            & MnR             \\
			\hline\hline
			MM   & 2.12          & 9.08           & 13.84          & 36.18          & 96.00          & 174.87          \\
				
			OT   & 2.85          & 9.48           & 14.94          & 37.61          & 94.00          & \textbf{172.77} \\
			
			PO   & \textbf{3.13} & \textbf{10.50} & \textbf{16.32} & \textbf{38.34} & \textbf{93.33} & 180.06          \\
			\hline			
		\end{tabular}
	}
	\caption{\label{tab:malta_results:heuristics}Results on heuristically annotated \model\  dataset.}
\end{table}
\begin{table}[t!]
	\resizebox{\linewidth}{!}{
		\begin{tabular}{|c|c|c|c|c|c|c| }	
			\multicolumn{7}{c}{Text to Video Retrieval}\\
			\hline
			Loss & R@1           & R@5            & R@10           & R@50           & MdR            & MnR            \\
			\hline\hline
					
			MM   & \textbf{4.73} & 15.02          & 24.69          & \textbf{62.96} & 33.00          & 47.15          \\
				
			OT   & 3.5           & 16.67          & 27.98          & 62.14          & 31.67          & 45.8           \\
			
			PO   & 4.12          & \textbf{17.49} & \textbf{29.42} & 61.93          & \textbf{31.00} & \textbf{45.74} \\
			\hline
					
			\multicolumn{7}{c}{Video to Text Retrieval}\\ \hline
			Loss & R@1           & R@5            & R@10           & R@50           & MdR            & MnR            \\
			\hline\hline
					
			MM   & 5.35          & \textbf{18.11} & 25.31          & 62.55          & 32.83          & 46.65          \\
				
			OT   & 5.76          & 16.05          & 26.75          & 67.49          & 29.17          & \textbf{41.95} \\
			
			PO   & \textbf{5.97} & 16.46          & \textbf{26.95} & \textbf{68.93} & \textbf{27.83} & 42.09          \\
			\hline			
		\end{tabular}
	}
	\caption{\label{tab:malta_results:multi-lingual-rudder}Results on the \model\ dataset that is  (multi-lingually) trained using audio from all the 4 languages. Multi-lingual \model uses audio experts from four different languages (Hindi, Marathi, Tamil, Telugu).}
\end{table}

\begin{table}[h!]
	
	\resizebox{\linewidth}{!}{
		\begin{tabular}{|c|c| c| c| c| c| c |}
			\hline
			Loss & \textbf{Training Pts} & R@1            & R@5            & R@10           & MdR           & MnR           \\ [0.5ex] 
			\hline\hline
			MM   & 100                   & 63.75          & 96.88          & 99.38          & 1.00          & 1.77          \\ 
			PO   & 100                   & \textbf{67.50} & \textbf{96.88} & \textbf{99.38} & \textbf{1.00} & \textbf{1.72} \\  \hline \hline
			MM   & 1000                  & 65.00          & 96.88          & 99.38          & 1.00          & 1.74          \\ 
			PO   & 1000                  & \textbf{65.63} & \textbf{97.5}  & \textbf{99.38} & \textbf{1.00} & \textbf{1.74} \\  \hline
		\end{tabular}
	}
	\caption{Empirical results on synthetically simulated data for 100 and 1000 training samples.}
	\label{table:synthetic_results}
\end{table}
We now discuss our results and present the underlying analysis. 
In Table~\ref{tab:malta_results}, we present our main results on the \model\ dataset using manually-derived annotations and heuristically-derived annotations. Further, in Table~\ref{tab:malta_results:multi-lingual-rudder}, we present the results on the multi-lingual \model\ setting wherein the textual captions are in Marathi but audio-experts corresponding to other languages are also used.  For both these losses, we sample a single negative example from each batch which partly explains the deterioration in performance compared to MM.
More details about these implementations are available in the dataset website. 

We observe that adding augmented supervision and invoking our partial order loss yields consistent improvements across all the evaluation metrics, in both Tables~\ref{tab:malta_results} and~\ref{tab:malta_results:multi-lingual-rudder}. We also observe that multi-lingual training significantly improves the retrieval performance. These numbers clearly validate our claim about the utility of the partial order loss in limited-data conditions. 
To further confirm the effectiveness of the method, we perform similar experiments with textual queries from languages other than Marathi, such as Tamil, Telugu and Hindi (a mix of unrelated languages). Table~\ref{tab:rudder-uni-lingual} tabulates performances when both the audio and the text corresponds to the same language, whereas Table~\ref{tab:rudder-individual-multi-lingual} tabulates results when the text is in Marathi but the audio features are derived from a different language.
Additionally, we observe that our results with heuristically annotated \model\ are fairly comparable to those obtained with the manually annotated \model, thus proving that one can achieve good performance with reasonable heuristics without having to undertake any additional human evaluations. 
\begin{table}[h!]
	\centering
	\begin{tabular}{|c|c|c|c|c|c|c|c|c|}
		\hline
		\textbf{} & \multicolumn{3}{c|}{T2V Retrieval} & \multicolumn{3}{c|}{V2T Retrieval} \\ \hline
		Loss    & R@5            & MdR            & MnR             & R@5            & MdR            & MnR             \\ \hline\hline
		MM(hin) & 9.39           & 68.17          & 108.98          & 9.96           & 67.67          & \textbf{107.32} \\ 
		PO(hin) & \textbf{10.45} & \textbf{63.5}  & \textbf{108.49} & \textbf{10.24} & \textbf{59.67} & 110.19          \\ \hline
		MM(tam) & \textbf{17.94} & 38.83          & 56.09           & 17.46          & 33.5           & \textbf{50.44}  \\ 
		PO(tam) & 17.62          & \textbf{28.5}  & \textbf{53.74}  & \textbf{19.68} & \textbf{32.83} & 50.57           \\ \hline
		MM(tel) & 11.57          & 56.00          & 85.25           & 11.93          & 53.17          & 81.99           \\
		PO(tel) & \textbf{11.93} & \textbf{52.83} & \textbf{83.99}  & \textbf{12.57} & \textbf{51.67} & \textbf{81.58 } \\ \hline
	\end{tabular}
	\caption{Results on \model\ dataset with audio and captions in the same language.} 
	\label{tab:rudder-uni-lingual}
\end{table}
\begin{table}[h!]
	\centering
	\begin{tabular}{|c|c|c|c|c|c|c|c|c|}
		\hline
		\textbf{} & \multicolumn{3}{c|}{T2V Retrieval} & \multicolumn{3}{c|}{V2T Retrieval} \\ \hline
		Loss    & R@5            & MdR             & MnR             & R@5            & MdR             & MnR             \\ \hline
		MM(hin) & 3.87           & 125.17          & 143.19          & 4.38           & 124.            & 142.99          \\ 
		PO(hin) & \textbf{4.21}  & \textbf{119.67} & \textbf{141.49} & \textbf{5.41}  & \textbf{122.33} & \textbf{142.93} \\ \hline
		MM(tam) & \textbf{10.02} & 54.67           & 56.41           & 8.86           & 45.5            & \textbf{52.94}  \\
		PO(tam) & 9.56           & \textbf{46.67}  & \textbf{56.38}  & \textbf{9.56}  & \textbf{43.67}  & 54.07           \\ \hline
		MM(tel) & 11.29          & 50.5            & 84.21           & 12.2           & 50.25           & 83.73           \\ 
		PO(tel) & \textbf{17.85} & \textbf{35.00}  & \textbf{68.09}  & \textbf{14.03} & \textbf{40.17}  & \textbf{70.81}  \\ \hline
	\end{tabular}
		
	\caption{Results on \model\ dataset with audio in different languages and captions in Marathi.}
	\label{tab:rudder-individual-multi-lingual}
	
\end{table}

\begin{table}[h!]
	\centering
	\resizebox{\linewidth}{!}{
		\begin{tabular}{|c|c|c|c|c|c|c|c|c|}
			\hline
			\textbf{} & \multicolumn{4}{c|}{T2V Retrieval} & \multicolumn{4}{c|}{V2T Retrieval} \\ \hline
			Loss               & R@1          & R@10          & MdR           & MnR            & R@1           & R@10          & MdR           & MnR           \\ \hline\hline
			MM~\cite{Liu2019a} & 4.8          & 25.0          & 43.3          & 183.1          & 8.4           & 37.1          & 20.3          & 87.2          \\ \hline
			MM                 & 5.6          & 27.6          & 35.3          & \textbf{153.7} & 10.5          & 41.5          & 16.0          & \textbf{69.5} \\		
			OT                 & 5.6          & 27.3          & 38.0          & 176.2          & 9.4           & 39.6          & 17.7          & 83.9          \\		
			PO                 & \textbf{6.2} & \textbf{28.8} & \textbf{33.3} & 161.29         & \textbf{11.2} & \textbf{43.4} & \textbf{14.7} & 74.1          \\		 \hline
		\end{tabular}
	}
	\caption{\label{tab:msrvtt_results}Results on MSRVTT dataset using 1 caption.} 
\end{table}

\paragraph{Additional Results:} We also demonstrate the performance of the proposed loss function on three benchmark datasets: MSR-VTT, DiDeMO and Charades-STA. MSR-VTT is a large-scale dataset with up to 20 captions per video. To conform better to our low-resource setting, we use only 1 caption per video\footnote{We also show results on the complete MSR-VTT dataset using all captions in the supplementary material. We do not see benefits from PO over the other losses in this setting given that evaluation using 20 captions has already a somewhat saturated coverage.}.
Table~\ref{tab:msrvtt_results} shows the results with our loss functions, along with the numbers reported in~\cite{Liu2019a} for this setting (MM~\cite{Liu2019a}). We observe that PO outperforms all other loss functions on all metrics (with the exception of MnR), and significantly improves over MM~\cite{Liu2019a}. Tables~\ref{tab:didemo_results} and~\ref{tab:charades_results} shows results on the DiDeMO and Charades-STA datasets, respectively. PO performs fairly well on the Charades-STA dataset. On DiDeMO, it improves upon the state-of-the-art on MdR and higher $k$ values for R@k. Interestingly, we observe a deviation from the trend for R@1 aznd R@5 for the V2T task. We believe this can be attributed to the captions in DiDeMO being of fairly large length on average, which hampers the ability of simple heuristics such as our noun-verb heuristic to identify useful partial samples%
\footnote{We note here that we only report baselines presented in prior work for each dataset; all baselines are not listed for each dataset. For Charades-STA, HN baseline results were too poor to even be reported.}.

\paragraph{Significance-test:} The significance of gains of PO over  baselines such as Triplet, DW, CE and HN are evident from the Tables~\ref{tab:malta_results}, \ref{tab:charades_results} and \ref{tab:didemo_results}. Hence we report  Wilcoxon's signed-rank test on the median ranks of PO and the tightly contending MM losses for all the experiments in the paper. We observe statistical significance at a \textit{p}-value less than $0.0001$ in favour of PO against MM. We also observe a \textit{p}-value less than $0.01$ for PO against OT.

\begin{table}[h!]
	\centering
	\resizebox{\linewidth}{!}{
		\begin{tabular}{|c|c|c|c|c|c|c|c|c|}
			\hline
			\textbf{} & \multicolumn{4}{c|}{T2V Retrieval} & \multicolumn{4}{c|}{V2T Retrieval} \\ \hline
			Loss              & R@1           & R@10          & MdR          & MnR           & R@1           & R@10          & MdR          & MnR           \\ \hline\hline
			S2VT              & 11.9          & -             & 13           & -             & 13.2          & -             & 15           & -             \\
			FSE               & 13.9          & -             & 11           & -             & 13.1          & -             & 12           & -             \\
			MM\cite{Liu2019a} & 16.1          & -             & 8.3          & 43.7          & 15.6          & -             & 8.2          & 42.4          \\
			HN                & 15.0          & 51.5          & 10.          & 47.6          & 14.1          & 51.1          & 10.          & 42.4          \\ \hline
			MM                & \textbf{16.4} & 54.5          & \textbf{8.0} & 44.4          & \textbf{17.0} & 52.3          & 10.0         & 42.6          \\
			OT                & 15.0          & 55.0          & 9.0          & 41.2          & 14.1          & 53.5          & 9.0          & \textbf{38.3} \\
			
			PO                & 16.3          & \textbf{56.5} & \textbf{8.0} & \textbf{40.2} & 15.0          & \textbf{54.9} & \textbf{8.0} & 39.6          \\
			\hline
		\end{tabular}
	}
	\caption{\label{tab:didemo_results}Results on DiDeMo dataset}
\end{table}

\begin{table}[h!]
	\centering
	\resizebox{\linewidth}{!}{
		\begin{tabular}{|c|c|c|c|c|c|c|c|c|}
			\hline
			\textbf{} & \multicolumn{4}{c|}{T2V Retrieval} & \multicolumn{4}{c|}{V2T Retrieval} \\ \hline
			Loss    & R@1          & R@10          & MdR           & MnR            & R@1          & R@10          & MdR           & MnR            \\ \hline
			Triplet & 1.6          & 9.2           & 153.0         & 240.5          & 1.3          & 8.0           & 157.2         & 252.0          \\
			DW      & 1.7          & 9.7           & 147.0         & 236.1          & 1.1          & 7.8           & 150.3         & 240.2          \\ \hline
			MM      & 2.4          & 14.3          & 89.0          & 174.6          & 2.0          & 13.4          & 86.8          & 174.4          \\				
			OT      & 2.7          & 15.5          & 80.5          & 171.4          & 2.0          & 14.2          & \textbf{82.3} & 172.3          \\		
			PO      & \textbf{3.6} & \textbf{15.9} & \textbf{77.0} & \textbf{162.3} & \textbf{3.2} & \textbf{14.9} & 83.0          & \textbf{164.6} \\ \hline
		\end{tabular} 
	}
	\caption{Results on Charades-STA dataset.}
	
	\label{tab:charades_results}
\end{table}
\section{Conclusion}
In this paper, we introduced a multi-lingual multi-modal video dataset, \model, that provides instructional videos of making toys from trash with both, audio and textual transcriptions, in six Indic languages. The dataset is the first of its kind and has the potential to further spur research in developing cross-linguo-modal embeddings, especially for under-represented languages. The dataset also comes with an evaluation set with multiple tracks for evaluating future works. We also introduced a novel partial-order contrastive loss that helps learn denser embeddings, thereby increasing the coverage across the embedding space. We present strong baselines on the \model dataset and demonstrate the effectiveness of the loss on other existing benchmarks for video-text retrieval as well. We hope the introduction of this resource would stimulate future works in the domain of cross-lingual video based Question-Answering, Video Summarization, Video-Text Retrieval, etc.

\bibliographystyle{ACM-Reference-Format}
\bibliography{references}  


\begin{thebibliography}{37}


\ifx \showCODEN    \undefined \def \showCODEN     #1{\unskip}     \fi
\ifx \showDOI      \undefined \def \showDOI       #1{#1}\fi
\ifx \showISBNx    \undefined \def \showISBNx     #1{\unskip}     \fi
\ifx \showISBNxiii \undefined \def \showISBNxiii  #1{\unskip}     \fi
\ifx \showISSN     \undefined \def \showISSN      #1{\unskip}     \fi
\ifx \showLCCN     \undefined \def \showLCCN      #1{\unskip}     \fi
\ifx \shownote     \undefined \def \shownote      #1{#1}          \fi
\ifx \showarticletitle \undefined \def \showarticletitle #1{#1}   \fi
\ifx \showURL      \undefined \def \showURL       {\relax}        \fi
\providecommand\bibfield[2]{#2}
\providecommand\bibinfo[2]{#2}
\providecommand\natexlab[1]{#1}
\providecommand\showeprint[2][]{arXiv:#2}

\bibitem[\protect\citeauthoryear{Arandjelovic, Gronat, Torii, Pajdla, and
  Sivic}{Arandjelovic et~al\mbox{.}}{2016}]%
        {arandjelovic2016netvlad}
\bibfield{author}{\bibinfo{person}{Relja Arandjelovic}, \bibinfo{person}{Petr
  Gronat}, \bibinfo{person}{Akihiko Torii}, \bibinfo{person}{Tomas Pajdla},
  {and} \bibinfo{person}{Josef Sivic}.} \bibinfo{year}{2016}\natexlab{}.
\newblock \showarticletitle{NetVLAD: CNN architecture for weakly supervised
  place recognition}. In \bibinfo{booktitle}{\emph{CVPR}}.
\newblock


\bibitem[\protect\citeauthoryear{{Bredin}, {Yin}, {Coria}, {Gelly},
  {Korshunov}, {Lavechin}, {Fustes}, {Titeux}, {Bouaziz}, and {Gill}}{{Bredin}
  et~al\mbox{.}}{2020}]%
        {VAD}
\bibfield{author}{\bibinfo{person}{Herv{\'e} {Bredin}},
  \bibinfo{person}{Ruiqing {Yin}}, \bibinfo{person}{Juan~Manuel {Coria}},
  \bibinfo{person}{Gregory {Gelly}}, \bibinfo{person}{Pavel {Korshunov}},
  \bibinfo{person}{Marvin {Lavechin}}, \bibinfo{person}{Diego {Fustes}},
  \bibinfo{person}{Hadrien {Titeux}}, \bibinfo{person}{Wassim {Bouaziz}}, {and}
  \bibinfo{person}{Marie-Philippe {Gill}}.} \bibinfo{year}{2020}\natexlab{}.
\newblock \showarticletitle{{pyannote.audio: neural building blocks for speaker
  diarization}}. In \bibinfo{booktitle}{\emph{ICASSP 2020, IEEE International
  Conference on Acoustics, Speech, and Signal Processing}}.
  \bibinfo{address}{Barcelona, Spain}.
\newblock


\bibitem[\protect\citeauthoryear{Carreira and Zisserman}{Carreira and
  Zisserman}{2017}]%
        {i3d}
\bibfield{author}{\bibinfo{person}{Joao Carreira} {and} \bibinfo{person}{Andrew
  Zisserman}.} \bibinfo{year}{2017}\natexlab{}.
\newblock \showarticletitle{Quo Vadis, Action Recognition? A New Model and the
  Kinetics Dataset}. In \bibinfo{booktitle}{\emph{CVPR}}.
\newblock


\bibitem[\protect\citeauthoryear{Chechik, Sharma, Shalit, and Bengio}{Chechik
  et~al\mbox{.}}{2010}]%
        {chechik2010large}
\bibfield{author}{\bibinfo{person}{Gal Chechik}, \bibinfo{person}{Varun
  Sharma}, \bibinfo{person}{Uri Shalit}, {and} \bibinfo{person}{Samy Bengio}.}
  \bibinfo{year}{2010}\natexlab{}.
\newblock \showarticletitle{Large Scale Online Learning of Image Similarity
  Through Ranking.}
\newblock \bibinfo{journal}{\emph{Journal of Machine Learning Research}}
  (\bibinfo{year}{2010}).
\newblock


\bibitem[\protect\citeauthoryear{Cuturi}{Cuturi}{2013}]%
        {cuturi2013sinkhorn}
\bibfield{author}{\bibinfo{person}{Marco Cuturi}.}
  \bibinfo{year}{2013}\natexlab{}.
\newblock \showarticletitle{Sinkhorn Distances: Lightspeed Computation of
  Optimal Transport}. In \bibinfo{booktitle}{\emph{Proceedings of the 26th
  International Conference on Neural Information Processing Systems - Volume
  2}} \emph{(\bibinfo{series}{NIPS’13})}. \bibinfo{publisher}{Curran
  Associates Inc.}, \bibinfo{address}{Red Hook, NY, USA},
  \bibinfo{pages}{2292–2300}.
\newblock


\bibitem[\protect\citeauthoryear{{Dong}, {Li}, and {Snoek}}{{Dong}
  et~al\mbox{.}}{2018a}]%
        {8353472}
\bibfield{author}{\bibinfo{person}{J. {Dong}}, \bibinfo{person}{X. {Li}}, {and}
  \bibinfo{person}{C.~G.~M. {Snoek}}.} \bibinfo{year}{2018}\natexlab{a}.
\newblock \showarticletitle{Predicting Visual Features From Text for Image and
  Video Caption Retrieval}.
\newblock \bibinfo{journal}{\emph{IEEE Transactions on Multimedia}}
  \bibinfo{volume}{20}, \bibinfo{number}{12} (\bibinfo{year}{2018}).
\newblock


\bibitem[\protect\citeauthoryear{{Dong}, {Li}, {Xu}, {Ji}, {He}, {Yang}, and
  {Wang}}{{Dong} et~al\mbox{.}}{2018b}]%
        {2018arXiv180906181D}
\bibfield{author}{\bibinfo{person}{Jianfeng {Dong}}, \bibinfo{person}{Xirong
  {Li}}, \bibinfo{person}{Chaoxi {Xu}}, \bibinfo{person}{Shouling {Ji}},
  \bibinfo{person}{Yuan {He}}, \bibinfo{person}{Gang {Yang}}, {and}
  \bibinfo{person}{Xun {Wang}}.} \bibinfo{year}{2018}\natexlab{b}.
\newblock \showarticletitle{{Dual Encoding for Zero-Example Video Retrieval}}.
\newblock \bibinfo{journal}{\emph{arXiv e-prints}} (\bibinfo{date}{Sept.}
  \bibinfo{year}{2018}), \bibinfo{pages}{arXiv:1809.06181}.
\newblock


\bibitem[\protect\citeauthoryear{Frome, Corrado, Shlens, Bengio, Dean, Ranzato,
  and Mikolov}{Frome et~al\mbox{.}}{2013}]%
        {im1}
\bibfield{author}{\bibinfo{person}{Andrea Frome}, \bibinfo{person}{Greg~S
  Corrado}, \bibinfo{person}{Jon Shlens}, \bibinfo{person}{Samy Bengio},
  \bibinfo{person}{Jeff Dean}, \bibinfo{person}{Marc\textquotesingle~Aurelio
  Ranzato}, {and} \bibinfo{person}{Tomas Mikolov}.}
  \bibinfo{year}{2013}\natexlab{}.
\newblock \showarticletitle{DeViSE: A Deep Visual-Semantic Embedding Model}.
\newblock In \bibinfo{booktitle}{\emph{NeurIPS}}.
\newblock


\bibitem[\protect\citeauthoryear{Gao, Mao, Zhou, Huang, Wang, and Xu}{Gao
  et~al\mbox{.}}{2015}]%
        {Gao2015AreYT}
\bibfield{author}{\bibinfo{person}{Haoyuan Gao}, \bibinfo{person}{Junhua Mao},
  \bibinfo{person}{Jie Zhou}, \bibinfo{person}{Zhiheng Huang},
  \bibinfo{person}{Lei Wang}, {and} \bibinfo{person}{Wei Xu}.}
  \bibinfo{year}{2015}\natexlab{}.
\newblock \showarticletitle{Are You Talking to a Machine? Dataset and Methods
  for Multilingual Image Question Answering}. In
  \bibinfo{booktitle}{\emph{Proceedings of the 29th Conference on Neural
  Information Processing Systems (NeurIPS)}}.
\newblock


\bibitem[\protect\citeauthoryear{Gao, Sun, Yang, and Nevatia}{Gao
  et~al\mbox{.}}{2017}]%
        {gao2017tall}
\bibfield{author}{\bibinfo{person}{Jiyang Gao}, \bibinfo{person}{Chen Sun},
  \bibinfo{person}{Zhenheng Yang}, {and} \bibinfo{person}{Ram Nevatia}.}
  \bibinfo{year}{2017}\natexlab{}.
\newblock \showarticletitle{Tall: Temporal activity localization via language
  query}. In \bibinfo{booktitle}{\emph{ICCV}}.
\newblock


\bibitem[\protect\citeauthoryear{Hadsell, Chopra, and LeCun}{Hadsell
  et~al\mbox{.}}{2006}]%
        {hadsell2006dimensionality}
\bibfield{author}{\bibinfo{person}{Raia Hadsell}, \bibinfo{person}{Sumit
  Chopra}, {and} \bibinfo{person}{Yann LeCun}.}
  \bibinfo{year}{2006}\natexlab{}.
\newblock \showarticletitle{Dimensionality reduction by learning an invariant
  mapping}. In \bibinfo{booktitle}{\emph{CVPR}}.
\newblock


\bibitem[\protect\citeauthoryear{Hahn, Silva, and Rehg}{Hahn
  et~al\mbox{.}}{2019}]%
        {meera2019action}
\bibfield{author}{\bibinfo{person}{Meera Hahn}, \bibinfo{person}{Andrew Silva},
  {and} \bibinfo{person}{James~M. Rehg}.} \bibinfo{year}{2019}\natexlab{}.
\newblock \showarticletitle{Action2Vec: A Crossmodal Embedding Approach to
  Action Learning}. In \bibinfo{booktitle}{\emph{CVPR-W}}.
\newblock


\bibitem[\protect\citeauthoryear{Hendricks, Wang, Shechtman, Sivic, Darrell,
  and Russell}{Hendricks et~al\mbox{.}}{2018}]%
        {hendricks-etal-2018-localizing}
\bibfield{author}{\bibinfo{person}{Lisa~Anne Hendricks},
  \bibinfo{person}{Oliver Wang}, \bibinfo{person}{Eli Shechtman},
  \bibinfo{person}{Josef Sivic}, \bibinfo{person}{Trevor Darrell}, {and}
  \bibinfo{person}{Bryan Russell}.} \bibinfo{year}{2018}\natexlab{}.
\newblock \showarticletitle{Localizing Moments in Video with Temporal
  Language}. In \bibinfo{booktitle}{\emph{EMNLP}}.
\newblock


\bibitem[\protect\citeauthoryear{Hershey, Chaudhuri, Ellis, Gemmeke, Jansen,
  Moore, Plakal, Platt, Saurous, Seybold, et~al\mbox{.}}{Hershey
  et~al\mbox{.}}{2017}]%
        {vggish}
\bibfield{author}{\bibinfo{person}{Shawn Hershey}, \bibinfo{person}{Sourish
  Chaudhuri}, \bibinfo{person}{Daniel~PW Ellis}, \bibinfo{person}{Jort~F
  Gemmeke}, \bibinfo{person}{Aren Jansen}, \bibinfo{person}{R~Channing Moore},
  \bibinfo{person}{Manoj Plakal}, \bibinfo{person}{Devin Platt},
  \bibinfo{person}{Rif~A Saurous}, \bibinfo{person}{Bryan Seybold},
  {et~al\mbox{.}}} \bibinfo{year}{2017}\natexlab{}.
\newblock \showarticletitle{CNN architectures for large-scale audio
  classification}. In \bibinfo{booktitle}{\emph{ICASSP}}.
\newblock


\bibitem[\protect\citeauthoryear{Hu, Shen, Albanie, Sun, and Wu}{Hu
  et~al\mbox{.}}{2020}]%
        {object_senet}
\bibfield{author}{\bibinfo{person}{Jie Hu}, \bibinfo{person}{Li Shen},
  \bibinfo{person}{Samuel Albanie}, \bibinfo{person}{Gang Sun}, {and}
  \bibinfo{person}{Enhua Wu}.} \bibinfo{year}{2020}\natexlab{}.
\newblock \showarticletitle{Squeeze-and-Excitation Networks}.
\newblock \bibinfo{journal}{\emph{IEEE TPAMI}} (\bibinfo{year}{2020}).
\newblock


\bibitem[\protect\citeauthoryear{Jaderberg, Simonyan, Vedaldi, and
  Zisserman}{Jaderberg et~al\mbox{.}}{2014}]%
        {scene}
\bibfield{author}{\bibinfo{person}{Max Jaderberg}, \bibinfo{person}{Karen
  Simonyan}, \bibinfo{person}{Andrea Vedaldi}, {and} \bibinfo{person}{Andrew
  Zisserman}.} \bibinfo{year}{2014}\natexlab{}.
\newblock \showarticletitle{Synthetic data and artificial neural networks for
  natural scene text recognition}.
\newblock \bibinfo{journal}{\emph{arXiv preprint arXiv:1406.2227}}
  (\bibinfo{year}{2014}).
\newblock


\bibitem[\protect\citeauthoryear{{Karaman}, {Gundogdu}, {Koç}, and
  {Alatan}}{{Karaman} et~al\mbox{.}}{2019}]%
        {Karaman_2019_IEEE}
\bibfield{author}{\bibinfo{person}{K. {Karaman}}, \bibinfo{person}{E.
  {Gundogdu}}, \bibinfo{person}{A. {Koç}}, {and} \bibinfo{person}{A.~A.
  {Alatan}}.} \bibinfo{year}{2019}\natexlab{}.
\newblock \showarticletitle{Quadruplet Selection Methods for Deep Embedding
  Learning}. In \bibinfo{booktitle}{\emph{ICIP}}.
\newblock


\bibitem[\protect\citeauthoryear{{Kiros}, {Salakhutdinov}, and {Zemel}}{{Kiros}
  et~al\mbox{.}}{2014}]%
        {im3}
\bibfield{author}{\bibinfo{person}{Ryan {Kiros}}, \bibinfo{person}{Ruslan
  {Salakhutdinov}}, {and} \bibinfo{person}{Richard~S. {Zemel}}.}
  \bibinfo{year}{2014}\natexlab{}.
\newblock \showarticletitle{{Unifying Visual-Semantic Embeddings with
  Multimodal Neural Language Models}}.
\newblock \bibinfo{journal}{\emph{arXiv e-prints}} (\bibinfo{date}{Nov.}
  \bibinfo{year}{2014}).
\newblock


\bibitem[\protect\citeauthoryear{Liu, Albanie, Nagrani, and Zisserman}{Liu
  et~al\mbox{.}}{2019}]%
        {Liu2019a}
\bibfield{author}{\bibinfo{person}{Y. Liu}, \bibinfo{person}{S. Albanie},
  \bibinfo{person}{A. Nagrani}, {and} \bibinfo{person}{A. Zisserman}.}
  \bibinfo{year}{2019}\natexlab{}.
\newblock \showarticletitle{Use What You Have: Video retrieval using
  representations from collaborative experts}. In
  \bibinfo{booktitle}{\emph{BMVC}}.
\newblock


\bibitem[\protect\citeauthoryear{Miech, Laptev, and Sivic}{Miech
  et~al\mbox{.}}{2019}]%
        {Antoine_icc19}
\bibfield{author}{\bibinfo{person}{Antoine Miech}, \bibinfo{person}{Ivan
  Laptev}, {and} \bibinfo{person}{Josef Sivic}.}
  \bibinfo{year}{2019}\natexlab{}.
\newblock \showarticletitle{Learning a Text-Video Embedding from Incomplete and
  Heterogeneous Data}. In \bibinfo{booktitle}{\emph{ICCV}}.
\newblock


\bibitem[\protect\citeauthoryear{Mithun, Li, Metze, and Roy-Chowdhury}{Mithun
  et~al\mbox{.}}{2018}]%
        {Mithun2018JointEW}
\bibfield{author}{\bibinfo{person}{Niluthpol~Chowdhury Mithun},
  \bibinfo{person}{Juncheng Li}, \bibinfo{person}{Florian Metze}, {and}
  \bibinfo{person}{Amit~K. Roy-Chowdhury}.} \bibinfo{year}{2018}\natexlab{}.
\newblock \showarticletitle{Joint embeddings with multimodal cues for
  video-text retrieval}.
\newblock \bibinfo{journal}{\emph{International Journal of Multimedia
  Information Retrieval}}  \bibinfo{volume}{8} (\bibinfo{year}{2018}),
  \bibinfo{pages}{3--18}.
\newblock


\bibitem[\protect\citeauthoryear{Rubner, Tomasi, and Guibas}{Rubner
  et~al\mbox{.}}{2000}]%
        {rubner2000earth}
\bibfield{author}{\bibinfo{person}{Yossi Rubner}, \bibinfo{person}{Carlo
  Tomasi}, {and} \bibinfo{person}{Leonidas~J Guibas}.}
  \bibinfo{year}{2000}\natexlab{}.
\newblock \showarticletitle{The earth mover's distance as a metric for image
  retrieval}.
\newblock \bibinfo{journal}{\emph{International Journal of Computer Vision}}
  (\bibinfo{year}{2000}).
\newblock


\bibitem[\protect\citeauthoryear{Sanabria, Caglayan, Palaskar, Elliott,
  Barrault, Specia, and Metze}{Sanabria et~al\mbox{.}}{2018}]%
        {How2}
\bibfield{author}{\bibinfo{person}{Ramon Sanabria}, \bibinfo{person}{Ozan
  Caglayan}, \bibinfo{person}{Shruti Palaskar}, \bibinfo{person}{Desmond
  Elliott}, \bibinfo{person}{Lo{\"{\i}}c Barrault}, \bibinfo{person}{Lucia
  Specia}, {and} \bibinfo{person}{Florian Metze}.}
  \bibinfo{year}{2018}\natexlab{}.
\newblock \showarticletitle{How2: {A} Large-scale Dataset for Multimodal
  Language Understanding}.
\newblock \bibinfo{journal}{\emph{CoRR}}  \bibinfo{volume}{abs/1811.00347}
  (\bibinfo{year}{2018}).
\newblock
\showeprint[arxiv]{1811.00347}
\urldef\tempurl%
\url{http://arxiv.org/abs/1811.00347}
\showURL{%
\tempurl}


\bibitem[\protect\citeauthoryear{{Schroff}, {Kalenichenko}, and
  {Philbin}}{{Schroff} et~al\mbox{.}}{2015}]%
        {facenet}
\bibfield{author}{\bibinfo{person}{F. {Schroff}}, \bibinfo{person}{D.
  {Kalenichenko}}, {and} \bibinfo{person}{J. {Philbin}}.}
  \bibinfo{year}{2015}\natexlab{}.
\newblock \showarticletitle{FaceNet: A unified embedding for face recognition
  and clustering}. In \bibinfo{booktitle}{\emph{CVPR}}.
\newblock


\bibitem[\protect\citeauthoryear{Socher, Karpathy, Le, Manning, and Ng}{Socher
  et~al\mbox{.}}{2014}]%
        {socher2014grounded}
\bibfield{author}{\bibinfo{person}{Richard Socher}, \bibinfo{person}{Andrej
  Karpathy}, \bibinfo{person}{Quoc~V Le}, \bibinfo{person}{Christopher~D
  Manning}, {and} \bibinfo{person}{Andrew~Y Ng}.}
  \bibinfo{year}{2014}\natexlab{}.
\newblock \showarticletitle{Grounded compositional semantics for finding and
  describing images with sentences}.
\newblock \bibinfo{journal}{\emph{Transactions of the Association for
  Computational Linguistics}} (\bibinfo{year}{2014}).
\newblock


\bibitem[\protect\citeauthoryear{{Tran}, {Wang}, {Torresani}, {Ray}, {LeCun},
  and {Paluri}}{{Tran} et~al\mbox{.}}{2018}]%
        {action_r2p1d}
\bibfield{author}{\bibinfo{person}{D. {Tran}}, \bibinfo{person}{H. {Wang}},
  \bibinfo{person}{L. {Torresani}}, \bibinfo{person}{J. {Ray}},
  \bibinfo{person}{Y. {LeCun}}, {and} \bibinfo{person}{M. {Paluri}}.}
  \bibinfo{year}{2018}\natexlab{}.
\newblock \showarticletitle{A Closer Look at Spatiotemporal Convolutions for
  Action Recognition}. In \bibinfo{booktitle}{\emph{CVPR}}.
  \bibinfo{pages}{6450--6459}.
\newblock


\bibitem[\protect\citeauthoryear{Vallender}{Vallender}{1974}]%
        {vallender1974calculation}
\bibfield{author}{\bibinfo{person}{SS Vallender}.}
  \bibinfo{year}{1974}\natexlab{}.
\newblock \showarticletitle{Calculation of the Wasserstein distance between
  probability distributions on the line}.
\newblock \bibinfo{journal}{\emph{Theory of Probability \& Its Applications}}
  (\bibinfo{year}{1974}).
\newblock


\bibitem[\protect\citeauthoryear{Venugopalan, Xu, Donahue, Rohrbach, Mooney,
  and Saenko}{Venugopalan et~al\mbox{.}}{2015}]%
        {venugopalan-etal-2015-translating}
\bibfield{author}{\bibinfo{person}{Subhashini Venugopalan},
  \bibinfo{person}{Huijuan Xu}, \bibinfo{person}{Jeff Donahue},
  \bibinfo{person}{Marcus Rohrbach}, \bibinfo{person}{Raymond Mooney}, {and}
  \bibinfo{person}{Kate Saenko}.} \bibinfo{year}{2015}\natexlab{}.
\newblock \showarticletitle{Translating Videos to Natural Language Using Deep
  Recurrent Neural Networks}. In \bibinfo{booktitle}{\emph{NAACL-HLT}}.
\newblock


\bibitem[\protect\citeauthoryear{Villani}{Villani}{2008}]%
        {villani2008optimal}
\bibfield{author}{\bibinfo{person}{C{\'e}dric Villani}.}
  \bibinfo{year}{2008}\natexlab{}.
\newblock \bibinfo{booktitle}{\emph{Optimal transport: old and new}}.
\newblock \bibinfo{publisher}{Springer Science \& Business Media}.
\newblock


\bibitem[\protect\citeauthoryear{{Wang}, {Li}, {Huang}, and {Lazebnik}}{{Wang}
  et~al\mbox{.}}{2019}]%
        {wang_neighborhood}
\bibfield{author}{\bibinfo{person}{L. {Wang}}, \bibinfo{person}{Y. {Li}},
  \bibinfo{person}{J. {Huang}}, {and} \bibinfo{person}{S. {Lazebnik}}.}
  \bibinfo{year}{2019}\natexlab{}.
\newblock \showarticletitle{Learning Two-Branch Neural Networks for Image-Text
  Matching Tasks}.
\newblock \bibinfo{journal}{\emph{TPAMI}} (\bibinfo{year}{2019}).
\newblock


\bibitem[\protect\citeauthoryear{Wang, Wu, Chen, Li, Wang, and Wang}{Wang
  et~al\mbox{.}}{2019}]%
        {VATEX}
\bibfield{author}{\bibinfo{person}{Xin Wang}, \bibinfo{person}{Jiawei Wu},
  \bibinfo{person}{Junkun Chen}, \bibinfo{person}{Lei Li},
  \bibinfo{person}{Yuan{-}Fang Wang}, {and} \bibinfo{person}{William~Yang
  Wang}.} \bibinfo{year}{2019}\natexlab{}.
\newblock \showarticletitle{{VATEX:} {A} Large-Scale, High-Quality Multilingual
  Dataset for Video-and-Language Research}.
\newblock \bibinfo{journal}{\emph{CoRR}}  \bibinfo{volume}{abs/1904.03493}
  (\bibinfo{year}{2019}).
\newblock
\showeprint[arxiv]{1904.03493}
\urldef\tempurl%
\url{http://arxiv.org/abs/1904.03493}
\showURL{%
\tempurl}


\bibitem[\protect\citeauthoryear{Wray, Larlus, Csurka, and Damen}{Wray
  et~al\mbox{.}}{2019}]%
        {Michael2019fine}
\bibfield{author}{\bibinfo{person}{Michael Wray}, \bibinfo{person}{Diane
  Larlus}, \bibinfo{person}{Gabriela Csurka}, {and} \bibinfo{person}{Dima
  Damen}.} \bibinfo{year}{2019}\natexlab{}.
\newblock \showarticletitle{Fine-Grained Action Retrieval Through Multiple
  Parts-of-Speech Embeddings}. In \bibinfo{booktitle}{\emph{CVPR}}.
  \bibinfo{pages}{450--459}.
\newblock


\bibitem[\protect\citeauthoryear{Wu, Manmatha, Smola, and Krahenbuhl}{Wu
  et~al\mbox{.}}{2017}]%
        {wu2017iccv}
\bibfield{author}{\bibinfo{person}{Chao-Yuan Wu}, \bibinfo{person}{R.
  Manmatha}, \bibinfo{person}{Alexander~J. Smola}, {and}
  \bibinfo{person}{Philipp Krahenbuhl}.} \bibinfo{year}{2017}\natexlab{}.
\newblock \showarticletitle{Sampling Matters in Deep Embedding Learning}. In
  \bibinfo{booktitle}{\emph{ICCV}}.
\newblock


\bibitem[\protect\citeauthoryear{{Xie}, {Girshick}, {Dollár}, {Tu}, and
  {He}}{{Xie} et~al\mbox{.}}{2017}]%
        {object_resnext}
\bibfield{author}{\bibinfo{person}{S. {Xie}}, \bibinfo{person}{R. {Girshick}},
  \bibinfo{person}{P. {Dollár}}, \bibinfo{person}{Z. {Tu}}, {and}
  \bibinfo{person}{K. {He}}.} \bibinfo{year}{2017}\natexlab{}.
\newblock \showarticletitle{Aggregated Residual Transformations for Deep Neural
  Networks}. In \bibinfo{booktitle}{\emph{2017 IEEE Conference on Computer
  Vision and Pattern Recognition (CVPR)}}. \bibinfo{pages}{5987--5995}.
\newblock


\bibitem[\protect\citeauthoryear{Xu, Mei, Yao, and Rui}{Xu
  et~al\mbox{.}}{2016}]%
        {xu2016msr}
\bibfield{author}{\bibinfo{person}{Jun Xu}, \bibinfo{person}{Tao Mei},
  \bibinfo{person}{Ting Yao}, {and} \bibinfo{person}{Yong Rui}.}
  \bibinfo{year}{2016}\natexlab{}.
\newblock \showarticletitle{MSR-VTT: A large video description dataset for
  bridging video and language}. In \bibinfo{booktitle}{\emph{CVPR}}.
\newblock


\bibitem[\protect\citeauthoryear{Xu, Sun, and Liu}{Xu et~al\mbox{.}}{2019}]%
        {xu2019learning}
\bibfield{author}{\bibinfo{person}{Lin Xu}, \bibinfo{person}{Han Sun}, {and}
  \bibinfo{person}{Yuai Liu}.} \bibinfo{year}{2019}\natexlab{}.
\newblock \showarticletitle{Learning with batch-wise optimal transport loss for
  3d shape recognition}. In \bibinfo{booktitle}{\emph{CVPR}}.
\newblock


\bibitem[\protect\citeauthoryear{Zhang, Hu, and Sha}{Zhang
  et~al\mbox{.}}{2018}]%
        {Zhang_2018_ECCV}
\bibfield{author}{\bibinfo{person}{Bowen Zhang}, \bibinfo{person}{Hexiang Hu},
  {and} \bibinfo{person}{Fei Sha}.} \bibinfo{year}{2018}\natexlab{}.
\newblock \showarticletitle{Cross-Modal and Hierarchical Modeling of Video and
  Text}. In \bibinfo{booktitle}{\emph{ECCV}}.
\newblock


\end{thebibliography}
\end{document}